\documentclass[12pt]{article}
\usepackage{amsmath}
\usepackage{graphicx}
\usepackage{subfigure} 
\usepackage{hyperref}

\begin{document}

\author{Ernst Trojan \and \textit{Moscow Institute of Physics and Technology} \and 
\textit{PO Box 3, Moscow, 125080, Russia}}
\title{Superluminal neutrino energy spectrum of OPERA and MINOS}
\maketitle

\begin{abstract}
We analyze the velocity dependence on energy of superluminal neutrino
recorded by the OPERA and MINOS collaborations and manage to approximate the
energy spectrum by a power law $E=p+C p^a$ where parameters must be taken
in the range $a=0.40\div 1.18$ and $C=1.5\times 10^{-5}\div 4.15\times 10^{-4}$ (momentum and energy are expressed in GeV). 
This rough estimation is constrained by the errors of measurements, and new experimental data are requested.  
\end{abstract}

\section{Introduction}

Neutrino was believed to be a massless fermion with energy 
\begin{equation}
E=pc  \label{u}
\end{equation}
and group velocity 
\begin{equation}
v=\frac{dE}{dp}=c  \label{v0}
\end{equation}
equal to the speed of light $c=1$ (in relativistic units). The modern theory
expects, however, that neutrino has finite mass \cite{M2011} 
\begin{equation}
m=m_\nu <0.28\,\mathrm{eV}  \label{mu}
\end{equation}
that implies deviation from the energy spectrum (\ref{u}) and velocity $%
v\neq c$. Recent experiments of the OPERA Collaboration \cite{OP} have
revealed suprluminal motion of neutrino. The time between signals 
\begin{equation}
t_0+\delta t=\frac Lv  \label{t}
\end{equation}
measured at the baseline $L=730$~km, was less than $t_0=L/c$, implying that
a small delay $\delta t$ was negative and the velocity of neutrino 
\begin{equation}
v-1=-\frac{\delta t}L  \label{d}
\end{equation}
was definitely above the speed of light. The velocity shift (\ref{d}) revealed
almost no dependence on energy, and at the average energy 
\begin{equation}
E=17\,\mathrm{GeV}  \label{e0}
\end{equation}
the time delay 
\begin{equation}
\delta t=-\mathbf{57.8}\,\pm 7.8\left[ stat.\right]
\,_{-5.9}^{+8.3}[sys.]\,\,\mathrm{ns}  \label{e99}
\end{equation}
corresponded to 
\begin{equation}
v-1=\left( \mathbf{2.37}\,\pm 0.32\left[ stat.\right]
\,_{-0.24}^{+0.34}[sys.]\mathit{\,}\right) \mathbf{\times }10^{-5}
\label{del}
\end{equation}
Superluminal neutrino was also observed by the MINOS Collaboration \cite{MI}
as well as in supernova explosion SN1987a \cite{SN87} .

This fact is a serious puzzle to the researchers. There is no lack of
hypotheses to explain it \cite{SLN1}. However, the value of superluminal
velocity (\ref{del}) imposes a severe constraint on the energy spectrum. The
energy dependence of $v$ was also explored by the OPERA collaboration \cite
{OP}, but it was not possible to warrant solid data because it was beyond
the accuracy of the measurement on account of large errors. However, the
energy dependence of $\left( v-1\right) $ on $E$ can contain very important
information that may allow to understand the nature of neutrino. In the
present paper we analyze the experimental data of the OPERA \cite{OP} and
MINOS \cite{MI} and try to establish the range of possible energy spectrum
of superluminal neutrino.

\section{Neutrino is not free tachyon}

The most natural and plain idea is to treat neutrino as a massive tachyon
whose energy spectrum 
\begin{equation}
E=\sqrt{p^2-m^2}  \label{tah}
\end{equation}
yields superluminal group velocity 
\begin{equation}
v=\frac{dE}{dp}=\frac p{\sqrt{p^2-m^2}}  \label{v0}
\end{equation}
estimated as
\begin{equation}
v\simeq 1+\frac 12\frac{m^2}{p^2}\simeq 1+\frac 12\frac{m^2}{E^2}
\label{v00}
\end{equation} 
for an ultra-relativistic particle ($E\simeq p\gg m$). 
It implies a tiny positive shift above the speed of light. However,
according to (\ref{mu}) and (\ref{v00}), we cannot get estimation greater
than 
\begin{equation}
v-1\simeq 10^{-22}  \label{v0a}
\end{equation}
at $E=17\,$GeV. Otherwise, we have to expect very large tachyon mass $m\simeq 120\,%
\mathrm{MeV}$ that contradicts to the expected upper bound (\ref{mu}). It is clear
that superluminal neutrino cannot be a free tachyon with the energy spectrum
(\ref{tah}), neither a free tardyon (ordinary particle) with the energy
spectrum $E=\sqrt{p^2+m^2}$.

However, it is not a problem of the theory because there are many sophisticated
arguments explaining the superluiminal velocity of neutrino \cite{SLN1}.  
Nevertheless, it is highly desirable to know the energy dependence of quantity 
\begin{equation}
f\left[ E\right] =v-1  \label{fe}
\end{equation}
that is equivalent to dependence on the momentum $f[p]$ for an ultra-relativistic particle. 
The knowledge of this dependence allows to restore the energy spectrum of neutrino 
\begin{equation}
E=p+\int f[p]dp  \label{sp}
\end{equation}
and test hypotheses of its nature.

\section{Dependence on energy}

The OPERA collaboration \cite{OP} has also obtained the following data at
various energy 
\begin{equation}
E=13.8\mathrm{\,GeV\qquad }\delta t=-\mathbf{54.7}\pm 18.4\left[
stat.\right] \,_{-6.9}^{+7.3}[sys.]\,\,\mathrm{ns}  \label{t1}
\end{equation}
\begin{equation}
E=28.2\,\mathrm{GeV\qquad }\delta t=-\mathbf{61.1}\pm 13\left[ stat.\right]
\,_{-6.9}^{+7.3}[sys.]\,\,\mathrm{ns}  \label{t2}
\end{equation}
\begin{equation}
E=40.7\mathrm{\,GeV\qquad }\delta t=-\mathbf{68.1}\pm 19.1\left[
stat.\right] \,_{-6.9}^{+7.3}[sys.]\,\,\mathrm{ns}  \label{t3}
\end{equation}
that together with (\ref{e0})-(\ref{e99}) can be described by
proportionality \cite{POW} 
\begin{equation}
\delta t\sim E^{a-1}  \label{v4}
\end{equation}
and, according to (\ref{d}) and (\ref{v4}), we get \cite{P2,P3} 
\begin{equation}
f\left[ E\right] =v-1=AE^{a-1}  \label{aaa}
\end{equation} 
that is equivalent to \begin{equation}
f\left[ p\right] =v-1=Ap^{a-1}  \label{aap}
\end{equation} 
for an ultra-relativistic particle. 
Let us develop this interpretation in detail.

According to (\ref{e99}) and (\ref{t1})-(\ref{t3}), parameter $a$ must lay
within the range 
\begin{equation}
a=0.40\div 2.08  \label{a}
\end{equation}
Then, approximation (\ref{aaa}) yields the observed velocity (\ref{del}) within the accuracy of measurement 
when 
\begin{equation}
A=\left( 0.09\div 16.6\right) \times 10^{-5}  \label{au}
\end{equation}
where $E$ in (\ref{aaa})  is expressed in GeV. 

The MINOS Collaboration \cite{MI} has recorded superluminal velocity 
\begin{equation}
v-1=\left( \mathbf{5.1}\,\pm 2.9[stat.+sys.]\right) \mathbf{\times }10^{-5}
\label{min}
\end{equation}
for the low energy neutrino with energy spectrum peaked at approximately $E=3\,\mathrm{GeV}$ 
with a long high-energy tail extending to $E=120\,\mathrm{%
GeV}$ and baseline $L=734$~km. The energy spectrum taken in the form (\ref{s}) 
can yield velocity (\ref{min}) within the accuracy of measurement when parameter $a$ is chosen in the range 
\begin{equation}
a=0.14\div 1.18  \label{ab}
\end{equation}
and 
\begin{equation}
A=1.81\div 20.6  \label{ak}
\end{equation} 
Velocity (\ref{min}) is not achieved at $E=3\,\mathrm{GeV}$ according to (\ref{s}) if $a$ and $A$ are beyond (\ref{ab}) and (\ref{ak}). 

Combining (\ref{a}) , (\ref{au}), (\ref{ab}) and (\ref{ak}) we have more strict estimation  
\begin{equation}
a=0.40\div 1.18  \label{a1}
\end{equation}
\begin{equation}
A=\left( 1.81\div 16.6\right) \times 10^{-5}  \label{a2}
\end{equation}
Substituting (\ref{aap}) in (\ref{sp}) we get the energy spectrum of ultra-relativistic neutrino 
\begin{equation}
E=p+Cp^a\qquad C=\frac Aa=1.5\times 10^{-5}\div 4.15\times 10^{-4}   \label{s}
\end{equation}
that satisfies both the OPERA and MINOS experiment. 
For example, choosing $a=1$ in (\ref{s}), we obtain velocity (\ref{del}) and (\ref{min}) within the accuracy of measurements 
when $A=\left( 2.2\,\div 3.03\right) {\times }10^{-5}$. 

\section{Conclusion}

Superluminal neutrino observed in experiments \cite{OP,MI} cannot be a free
tachyon because it mass (\ref{mu}) is not enough to correspond the observed
velocity (\ref{del}).

The dependence (\ref{fe}) of the neutrino velocity $v$ on the energy $E$ may
give much information about its energy spectrum (\ref{sp}). It can be taken
as a power law (\ref{s}) with parameters (\ref{a1}) and (\ref{a2}). It
should be emphasized that parameter $a$ in (\ref{s}) must be positive, and
the real energy spectrum lays somewhere between $E=p+1.5\times
10^{-5}p^{1.18}$ and $E=p+4.15\times 10^{-4}p^{0.4}$. There is no
possibility to establish $a$ and $A$ at high accuracy because of errors of
experimental data. 
It is impossible even to clarify whether $a>1$ or $a<1$ and whether function $v[E]$ is monotonically increasing. 
Of course, new measurements will reveal the exact energy spectrum of neutrino and clarify its physical nature. 
Now we can only state that superluminal neutrino is not a free tachyon and that the second term in
the right side of (\ref{s}) may give a hint to nonlinear self-interaction or external field acting on neutrino. 
It is the subject of further theoretical work.

The author is grateful to Erwin Schmidt for discussions.

\end{document}